\documentstyle[12pt,aasms4]{article}
\received{}
\accepted{}
\journalid{337}{}
\articleid{11}{}

\slugcomment{Version date: August 31, 1999}

\lefthead{Weintraub et al.}
\righthead{Polarimetric Imaging of the Egg Nebula with NICMOS}

\begin{document}

\def\arcsec{{$^{\prime\prime}$}}

\title{Pinpointing the Position of the Post-AGB Star 
at the Core of RAFGL 2688 using Polarimetric Imaging with NICMOS}

\author{David A. Weintraub\altaffilmark{1}, Joel H. 
Kastner\altaffilmark{2},  Dean C. Hines\altaffilmark{3}, 
Raghvendra Sahai\altaffilmark{4}}

\altaffiltext{1}{Department of Physics \& Astronomy,
Vanderbilt University, P.O. Box 1807 Station B, Nashville, TN 37235; 
david.weintraub@vanderbilt.edu}

\altaffiltext{2}{Carlson Center for Imaging Science, RIT, 
       84 Lomb Memorial Drive, Rochester, NY 14623; jhk@juggler.mit.edu}

\altaffiltext{3}{Steward Observatory, The University of Arizona, Tucson, 
AZ 85721; dhines@as.arizona.edu}
 
\altaffiltext{4}{Jet Propulsion Laboratory, MS 183-900, California 
Institute of Technology, Pasadena, CA 91109; sahai@grandpa.jpl.nasa.gov}

\begin{abstract}

We have used infrared polarimetric imaging with NICMOS to determine
precisely the position of the star that illuminates (and presumably 
generated) the bipolar, pre-planetary reflection nebula RAFGL 2688 
(the Egg Nebula). The polarimetric data pinpoint the illuminating 
star, which is not detected directly at wavelengths $\le$ 2 
$\mu$m, at a position well within the dark lane that bisects the 
nebula, 0\farcs55 ($\sim550$ AU) southwest of the infrared peak 
which was previously detected at the southern tip of the northern 
polar lobe. The inferred position of the central star corresponds 
to the geometric center of the tips of the four principle lobes of 
near-infrared H$_2$ emission; identifying the central star at this
position also reveals the strong point symmetric structure of the
nebula, as seen both in the intensity and polarization structure 
of the polar lobes. The polarimetric and imaging data 
indicate that the infrared peak directly detected in the NICMOS 
images is a self-luminous source and, therefore, is most likely a 
distant binary companion to the illuminating star. Although present 
theory predicts that bipolar structure in pre-planetary and planetary 
nebulae is a consequence of binary star evolution, the separation between 
the components of the RAFGL 2688 binary system, as deduced from these 
observations, is much too large for the presence of the infrared 
companion to have influenced the structure of the RAFGL 2688 nebula.

\end{abstract}


\section{Introduction}

The bipolar structures exhibited by a substantial fraction of the known 
planetary nebulae likely arise during the last, rapid, pre-planetary 
nebula (PPN) stage of evolution of intermediate-mass (1--8 M$_\odot$) 
stars off the asymptotic giant branch (AGB). 
A popular, albeit largely untested, model for 
such bipolarity is that the central AGB star possesses a companion that 
aids in the buildup of a dense, dusty equatorial torus surrounding the 
central star (e.g., \cite{soke1998}).  Alternatively, the fossil remnant 
of a $\beta$ Pic-like main-sequence disk may bear responsibility for 
triggering bipolarity during post-main sequence evolution (\cite{kast1995}).  
Whatever the mechanism that abets their formation, bipolar PPN 
typically show two bright reflection lobes separated by a dark 
dust lane.  The star that illuminates the polar lobes presumably is located
at or near the center of the equatorial, dust torus. While this geometry
obscures the central star along our direct line of sight, photons readily
escape the nebular core in the polar directions and subsequently are 
scattered by dust grains located primarily in the walls of the rarefied, 
expanding lobes.  As even the lobe walls tend to be optically thin
in the near-infrared, such photons can be singly scattered out of the nebula
into our line of sight.  Single scattering produces polarized light that 
contains a record of the original direction of the unpolarized 
light source; therefore, polarimetric maps of such polarized nebulae contain
clues as to the locations of their illuminating sources, even if those 
stars lie hidden inside dust lanes. 

Recent direct imaging of RAFGL 2688 (the Egg Nebula) with the Near Infrared 
Camera and Multi-Object Spectrometer (NICMOS) aboard the Hubble Space 
Telescope ({\it HST}) (\cite{saha1998}) revealed a compact red source 
just south of the bottom of the northern reflection lobe.  However, 
initial analysis of the polarimetric maps from NICMOS indicated that 
this red source was not the primary illuminator of the reflection 
nebulosity; this object is most likely a companion to the post-AGB star that
lurks in the core of the Egg Nebula.  From a preliminary examination of 
the 2.0 $\mu$m polarimetric map, Sahai et al. suggested that the obscured, 
post-AGB star was located $\simeq$ 750 AU (0\farcs75) south of the red companion.

In this paper, we present a rigorous analysis of the 2.0 $\mu$m 
polarization map of RAFGL 2688 obtained by NICMOS.  We determine
the precise position of the post-AGB star in the core, assess the 
relationship of the red source to the illuminator star, and discuss
the implications of this work for understanding the formation 
of the Egg Nebula and of other bipolar PPN.

\section{Polarization Data Analysis}

The data and data reduction methods used in this study were first 
presented by Sahai et al.\ (1998). In brief summary, RAFGL 2688 was imaged
through the POL0L, POL120L and POL240L filters with camera 2 (NIC2) of NICMOS, using
integration times of 1215 s for each filter. These filters are centered
at 1.994 $\mu$m and have a full-width-half-maximum of 0.2025 $\mu$m.  
The field of view for these images is 19\farcs5$\times$19\farcs3 and 
the plate scale is 0\farcs076/pixel (\cite{thom1998}).  The calculations of 
fractional polarization ({\it p}) and polarization position angle 
($\theta$) are carried out as described by Hines (1998)\footnote{Note that the
coefficients for polarimetric imaging calculations have been updated; see 
http://www.stsci.edu/instruments/nicmos/nicmos\_polar.html and Hines, 
Schmidt \& Schneider 1999}; however,
we find that the best position angle calculations include the 
addition of a small, constant angle $\phi$ to $\theta$, i.e.
$$ \theta = {1 \over 2}\tan^{-1}\biggr({U \over Q}\biggl) + \phi,$$  
where $U$ and $Q$ are the Stokes vectors obtained from the polarimetric
images.  The offset angle $\phi$ could represent a systematic rotation of the filters in 
the polarization filter set from their nominal position angles.  For example, 
if the three polarizing filters were designed to lie at position angles 
0$^\circ$, 120$^\circ$, and 240$^\circ$, they actually 
are found at position angles 0$^\circ$ + $\phi$, 120$^\circ$ + $\phi$, 
and 240$^\circ$ + $\phi$.   Alternatively, $\phi$ could represent 
uncertainties in our knowledge of the absolute position angles assumed 
for the polarization calibrators.  We suggest that RAFGL 2688 represents 
the best absolute position angle calibrator for NICMOS polarimetric data.  
As explained in \S 4.1, we have determined empirically that 
$\phi$ = 4.0$^\circ$ $\pm$ 0.2$^\circ$.  
A polarization map of RAFGL 2688, made with $\phi$ = 4.0$^\circ$, 
is presented in Figure~1.  

\section{The Polarization Structure of the Nebula}

A centrosymmetric pattern is the dominant single feature of
the polarization map (Fig.~1); however, it is apparent by careful 
inspection of Fig.~1 that the polarimetric centroid is not spatially 
coincident with the source (labeled A) at the southern tip of the 
northern lobe (see \S 4).  
Overall, the nebula is very highly polarized, with virtually
the entire southern lobe polarized with $p$ $>$ 0.50 (see Fig.~6 in 
Sahai et al.\ 1998 for a grayscale map of the polarized intensity).
A second strong feature of the polarization structure is the 
apparent point symmetry of the polarization pattern around 
position B, which we describe below. 
The implication of such a symmetry for 
the origin of the bipolar lobes is discussed later (see \S 4.2).

The southern lobe is more highly polarized overall than the northern 
lobe (Figure~2).  In the north, only 11 pixels show vectors with 
polarization amplitudes above 0.7; all of these  vectors are on or  
west of the polar axis, with all but one at least 5\arcsec\ from the 
center of the nebula (Figure~2a). In contrast, $\sim$300 pixels in 
the southern lobe have $p$ $>$ 0.7; these pixels are dominantly on 
the eastern side of the polar axis and all of them lie more than 
4\arcsec\ from the center of the nebula, demonstrating a strong point 
symmetry to the polarization pattern around the nebular core.  
An additional $\sim$1000 pixels are polarized with 0.6 $<$ $p$ $<$ 0.7 
(Figure~2b).  In the north, virtually all these vectors lie west of 
the polar axis, stretching inwards along the west limb of the 
reflection lobe from a distance of $\sim$7\arcsec\ to just more than 
3\arcsec\ from the center.  In the south, these vectors are uniformly
spread across the lobe in the outer regions and more concentrated to
the east of the polar axis closer to the core.  Most 
of the rest of the southern lobe is polarized at a level $p$ $>$ 0.50 
(Figure~2c).  In the north, the polarization vectors in the range  
0.4 $<$ $p$ $<$ 0.6 cover most of the center of 
the lobe (Figure~2c); the region covered by these vectors 
stretches radially away from the core along the eastern side;
the polarization vectors in the range 0.4 $<$ $p$ $<$ 0.6  
also cover the center of the southern lobe at small radial distances 
and then this region stretches outwards from the core along the 
western side. Finally, the outer edges of the northern lobe nearest 
to the nebular core are dominated by polarization amplitudes in the 
0.15--0.40 range (Figure~2d).

\section{The Polarimetric Centroid}

\subsection{Method of Determination}

To determine the position of the source that illuminates the nebula,
we have used the method presented by Weintraub \& Kastner (1993),
coded into a program in the software package IDL.
This method takes advantage of the fact that a dust grain that
singly scatters photons out of the nebula imparts a polarization
position angle to the scattered light that is perpendicular to
the scattering plane, i.e., perpendicular to the projected direction 
from that dust grain to the source of  
illumination.  Thus, for every pair of polarization vectors in 
a map, we can draw perpendiculars to each vector and determine a
point of intersection.  Ideally, for
noiseless data and purely singly scattered photons, all the pairs
of vectors would have a unique intersection, the {\it polarimetric 
centroid}, which should mark the intersection between the polar axis 
and the disk midplane (assuming the illuminating source is identical 
with the central star of the nebula and that the central star lies 
at the geometric center of the nebula).  

Even for noisy data and a mixture of singly and multiply scattered
photons, one can use the method of intersections of polarization
perpendiculars to determine the polarization centroid,
albeit with finite positional error bars (\cite{wein1993}).  For a given data set,
the accuracy with which we can determine the centroid depends on the
absolute calibration of the position angles and thus depends on 
our knowledge of $\phi$. If $\phi$ is marginally inaccurate, the 
polarimetric centroid will be poorly determined while if $\phi$ is quite 
inaccurate, there will be no polarimetric centroid in the map at all. 
Thus, we have determined $\phi$ by examining a range of $\phi$ values 
between $-10^\circ$ and $+10^\circ$ and adopting the value 
that minimizes the uncertainty in determining the polarimetric centroid.
In calculating the polarimetric centroid, we limit the calculation to 
the $>$8000 pixels containing flux levels with signal-to-noise ratios 
greater than six in all three of the POL0L, POL120L, and POL240L images.

Many of these pairs of vectors have nearly parallel position angles. 
For vector pairs with similar position angles, especially given even a small 
error in determining the true position angles, the intersection position
is poorly determined.  We therefore impose an additional constraint: we
reject all vector pairs for which the angle between the vectors 
(modulo 180$^\circ$) is less than 20$^\circ$.  This ensures that the small 
uncertainties in the position angle calculations do not produce large 
uncertainties in the actual position of the centroid.  
In practice, in addition to noise, many of the pixels, usually those 
with polarization vectors with lower polarization amplitudes, represent 
parts of the reflection nebula in which multiple scattering is 
probably dominant.   Thus, for our final calculations, we placed a limit 
on the minimum allowable fractional polarization to be $p_{min}$ $\ge$ 0.15 
in order to exclude lines of sight dominated by multiple scattering. 

After calculating the intersection points for the complete set of 
allowable vectors and vector pairs, we calculate the statistical mean 
and the standard deviation of the mean ($\sigma$) for the polarization 
centroid.  We then repeat this calculation, keeping only intersection 
points within a 3-$\sigma$ rejection threshold of the initially 
determined mean. We continue with this process, iteratively, until the 
solution converges on the polarimetric centroid (denoted B).  
We find that the initial calculation typically lies 
within 0.1 pixels ($<$ 0\farcs01) of the final position and the calculation 
converges after only $\sim$five iterations and after rejecting only 
$\sim$2\%--4\% of the total possible intersections.  Changing the rejection 
threshold appears to affect only the size of the uncertainty and the rate of 
convergence, not the position of the polarization centroid itself.

\subsection{Results}

In Figure~3, we present the same map as shown in Figure~1 
but drawn with all the vectors perpendicular to the polarization position 
angles.  These vectors clearly point to a single intersection point, 
the polarimetric centroid (labeled B in Fig.~1,3-6).  In addition, this 
map illustrates, very 
clearly, the symmetry axis of the nebula, as seen in scattered light.

By examining solutions where $p_{min}$ ranges from 0.15 to 0.35, we 
find that the centroid lies 0\farcs52 $\pm$ 0\farcs02 west and 0\farcs16 
$\pm$ 0\farcs03 south (Figure 4) of the isolated intensity peak at the 
southern tip of the north lobe (position A), well within the dark dust 
lane that cuts across the middle of the bipolar nebula.  The positional 
uncertainty is dominated by the systematic differences between solutions found
when selecting different values of $p_{min}$, rather than by the statistical
errors in a single calculation (which are more than an order of magnitude 
smaller).  

We have used the position of the polarimetric centroid combined with 
the vector pattern to determine the direction of the projection of the 
polar (major) axis of the Egg Nebula.  One can see (Fig.~3) that the
projected polar axis, drawn at a position angle of 12$^\circ$ (east 
of north), runs exactly 
parallel to the straight lines formed by the alignment of the 
perpendiculars (of the polarization vectors) along the central axis of both the
north and south scattering lobes. A change in more than 1$^\circ$ 
in the position angle of the polar axis produces a clear error in the left-right 
symmetry of the lobes, as defined by the polarization vectors.
Thus, we believe this determination of the projected position 
angle of the polar axis represents an improvement over 
the previously inferred angle of 15$^\circ$ (\cite{ney1975}).

In projection, the centroid is located 
much closer to the northern than the southern lobe. The fact that B lies 
closer to the southern tip of the northern lobe than to the northern tip 
of the southern lobe is consistent with previous determinations that the 
polar axis of the system is inclined such that the northern lobe is 
tilted toward the observer.  This geometry causes the optically 
thick equatorial torus to obscure the innermost part of the southern 
lobe but permits us to view most of the inner regions of the northern lobe.

It is interesting to note the point symmetry between the two 
scattering lobes.  In the north, the majority of the total intensity 
of the nebula is east of the polar axis, including the brightest reflection
peaks (see Fig.~1).  In contrast, in the south, most of the reflection 
nebula is found to the west of the polar axis.  In both lobes, the 
morphologically larger side of the nebula is the side showing
lower overall polarization levels.  We also see that the polar
axis runs through the eastern side of the inward extension of the
southern lobe and through the western side of the inward extension
of the northern lobe.  The simplest mechanism for producing 
point symmetric structure in the nebula is the operation of collimated 
bipolar outflows. Sahai \& Trauger (1998) have argued, based on finding 
a high degree of point symmetry in the morphologies of their sample of 
young planetary nebulae, that such outflows are the primary agent for 
producing aspherical structure in planetary nebulae. 

\section{The Illuminator Star and its Surroundings}

The polarimetric centroid presumably marks the position of the 
post-AGB star that illuminates most or all of both the northern 
and southern reflection lobes of the Egg Nebula. We now consider 
whether this illuminator and the intensity peak A constitute a 
widely spaced ($>$550 AU) binary system.  

If a field star were at position A, such a star would reveal
itself in an Airy pattern in the total intensity profile, as NICMOS 
generates such patterns even for very faint point sources.  The 
absence of such a pattern indicates that the intensity peak A 
is an extended object.  Such an object could be either a region 
of enhanced dust density that reflects light from B or a star 
embedded in the nebula that illuminates and heats the local pocket 
of dust around it.  

In Figure~5, we present a polarization map of the same region as 
seen in Figure~4; however, in order to focus on the polarization 
behavior near A, we present in Fig.~5 only the polarization 
vectors with amplitudes $p$ $<$ 0.15.  If a point source at 
position A suffers little local extinction, then it becomes a 
source of 2 $\mu$m photons which should generate some sort of 
centrosymmetric polarization pattern centered on A while the direct 
line of sight to A should show a low polarization level.  Given the 
local presence of the illuminator star at B, we might expect this 
pattern to be distorted by the influence of a second photon source.  

In examining Fig.~5, we find neither an indication of any kind of 
centrosymmetric pattern, even a strongly distorted one, centered on 
the position of the intensity peak at A, nor a simple, 
centrosymmetric pattern focused on the position of the illuminator at B, 
similar to that which characterizes the vectors in the rest of the 
nebula.  Instead, close to A, we find 
a region marked by extremely low polarization levels and a 
disorganized polarization pattern, despite the fact that the 
signal-to-noise ratio is high.  A somewhat more 
organized vector pattern is seen in the vectors that 
lie northeast, north and northwest of A, and which appear
to define a centrosymmetric pattern centered on B.

If intensity peak A were simply a region of enhanced density of cold dust, 
we should see a pattern of highly polarized vectors at A suggesting 
direct illumination from position B, as is seen at other intermediate 
intensity peaks further out in the northern lobe.  The absence of 
such a pattern suggests that peak A is self-luminous; however, 
the lack of any Airy profile as would be expected from a point source 
indicates that the source at A, at 2 $\mu$m, is
seen as a small, extended nebula.  At this position in the 
nebula, the local NICMOS point spread function generated by emission 
from the extended source at A, combined with the illumination of dust 
in this vicinity by B, generates a disorganized polarization pattern 
marked by relatively low polarization levels.  This analysis therefore
supports the suggestion 
that intensity peak A is a self-luminous near-infrared source.

What is the nature of the self-luminous source at peak A?  Is it a deeply
embedded star or a blob of warm dust?  If it is a blob of warm dust, the 
only likely heat sources would be illumination from the former AGB star 
located at least 550 AU distant or shock heating.  To produce significant thermal 
emission at 1.65 $\mu$m, the wavelength at which the blob begins to appear
(\cite{saha1998}), would require dust with temperatures of at least 1000 K.
The heating of a large amount of dust when the heat source is at least 
550 AU away is highly unlikely, even for an AGB star with a luminosity of 
10$^4$ L$_\odot$.  In addition, some of the luminosity of the AGB star 
would be expected to show up in a reflection pattern at peak A, which we do not 
see. As for shock heating, the maps of H$_2$ 
emission (Fig.\ 6; also, see \cite{saha1998}) reveal no evidence of 
shocked gas within a few tenths of an arcsec (several hundred AU) of
peak A. If the dust had been heated by a passing shock that is now
200 AU away, having moved past at 30 km s$^{-1}$, it would have had
at least 30 years to cool down.  Thus, it appears more likely that
intensity peak A is a star and that A and B most likely constitute a 
widely spaced, binary star system.

Assuming A and B are a binary, their minimum separation is 550 AU (taking
d = 1 kpc).  If A and B are both in the equatorial plane and the polar 
axis is tilted 15$^\circ$ out of the plane of the sky 
(Sahai et al.\ (1998) estimated a tilt
of 10$^\circ$--20$^\circ$ from the axial ratio of the dust torus), 
then the star at 
A would lie $\sim$900 AU more distant than the star at B, making the 
true binary separation about 1000 AU.  This separation is several orders 
of magnitude larger than that hypothesized (\cite{morr1987,soke1998}) 
for a central binary system that could trigger the formation of an 
equatorial disk and the consequent bipolar outflow.  

It is remarkable that position B appears 
to be equidistant and point-symmetrically placed between 
the apex of the western loop (E1), the apex of the middle of the 
eastern loops (E3), and the most distant points in the polar 
lobes of molecular hydrogen emission (Fig.\ 6).  Thus, the 
polarimetric and molecular hydrogen emission centroids are 
positionally coincident.  This result strongly indicates that 
the nebular illuminator at B also generated the H$_2$ emission,
where the H$_2$ emission regions are delineated by sharp outer 
boundaries suggestive of shocks.  As shocks require fairly 
sudden changes --- in this case, perhaps the rapid turning-on of a 
fast wind from the former AGB star, perhaps triggered by the quite 
quick stripping and ejection of the stellar envelope and the 
subsequent capture of a close companion --- the relationship between 
position B and the H$_2$ emission lobes suggests that the shocks seen 
in the H$_2$ were caused by a very sudden event or series of events 
in the evolution of the central star. 

Thus, while the presence of the A+B binary at the core of RAFGL 2688 does 
not lend support to the binary trigger hypothesis for the formation 
of bipolar planetary nebulae, the relationship between the central star
at B and the H$_2$ lobes may support such a hypothesis.  
Specifically, absorption of a close binary companion by the atmosphere 
of the central AGB star may cause the ejection of high-velocity material;
the ejected material produces the shocked H$_2$ emission and 
generates the bipolar structure of the Egg Nebula.

\section{Summary}

From a detailed analysis of the polarimetric images obtained using 
NICMOS and the {\it HST}, we have precisely determined the position 
of the post-AGB star in the waist of the Egg Nebula and the projected 
orientation of the polar axis (PA 12$^\circ$) of this bipolar system. 
This post-AGB star, which illuminates the Egg Nebula, falls 
point-symmetrically at the center of the molecular hydrogen emission 
regions that mark the waist and the polar lobes of the nebula.  
We find that this star lies 550 AU in projected distance, and perhaps 
1000 AU in physical distance, from the star previously identified 
(\cite{saha1998}) at the southern tip of the northern polar lobe. 
Thus, these data provide clear evidence 
for the presence of an optically obscured, widely spaced binary system 
near the core of the bipolar, pre-planetary nebula RAFGL 2688.  However,
the separation between these components is orders of magnitude larger 
than required by models postulating that companions to AGB stars 
trigger the production of bipolar planetary nebulae.

\acknowledgments{
DCH acknowledges support by NASA grant NAG5-3042 to the 
NICMOS instrument definition team.  RS thanks NASA for support through 
grant GO-07423.01-96A from the Space Telescope Science Institute (which 
is operated by the Association of Universities for Research in Astronomy, 
Inc., under NASA contract NAS5-26555).
}

\clearpage

\figcaption[fig1.eps]{
Polarization map of RAFGL 2688, obtained from 2.0 $\mu$m imaging from
NICMOS.  Vectors are plotted only where intensity level is greater 
than 3-$\sigma$ in all three Stokes images. The polarimetric centroid 
is labeled B and found at the intersection of the lines marking the 
projected polar axis (12$^\circ$ east of north) and equatorial
plane of the nebula. The intensity peak is labeled A. Vectors indicate
the polarization strength (length) and position angles in each pixel,
with only vectors with $p$ $\ge$ 0.15 plotted.  
% Pixel scale is 0\farcs076.
In this and all other figures, orientation is indicated by north (N) 
and west (W) axes, offset distances are measured from 
B, absolute polarization amplitudes are indicated by a $p$ = 0.30
fiducial vector marked in the lower right corner of panel, and intensity
contours are drawn at 1 magnitude intervals, with the lowest contour at
the 3-$\sigma$ level in the total intensity image.
\label{fig1}}

\figcaption[fig2.eps]{Full view of RAFGL 2688 with polarization vectors 
in the specified range overlaid on 2.0 $\mu$m intensity contours (same
contour levels as Fig.\ 1). 
a) $p$ $>$ 0.70,
b) 0.70 $>$ $p$ $>$ 0.60, 
c) 0.60 $>$ $p$ $>$ 0.40, 
d) 0.40 $>$ $p$ $>$ 0.15.  
\label{fig2}}

\figcaption[fig3.eps]{Same as Figure~1 but with all polarization vectors drawn 
perpendicular to their normal orientations.   \label{fig3}}

\figcaption[fig4.eps]{Polarization map of central region of nebula 
showing position of polarimetric centroid. Only vectors with $p$ 
$\ge$ 0.15 are plotted.  Size of small plus sign at the position of
Source B indicates 1-$\sigma$ 
systematic uncertainty in determination of centroid position.  \label{fig4}}

\figcaption[fig5.eps]{Close-up polarization map of same region as Figure~4, 
but showing only vectors with $p$ $<$ 0.15.  
   \label{fig5}}

\figcaption[fig6.eps]{Molecular hydrogen emission (grayscale) overlaid 
with 2 $\mu$m continuum (contours).
\label{fig6}}

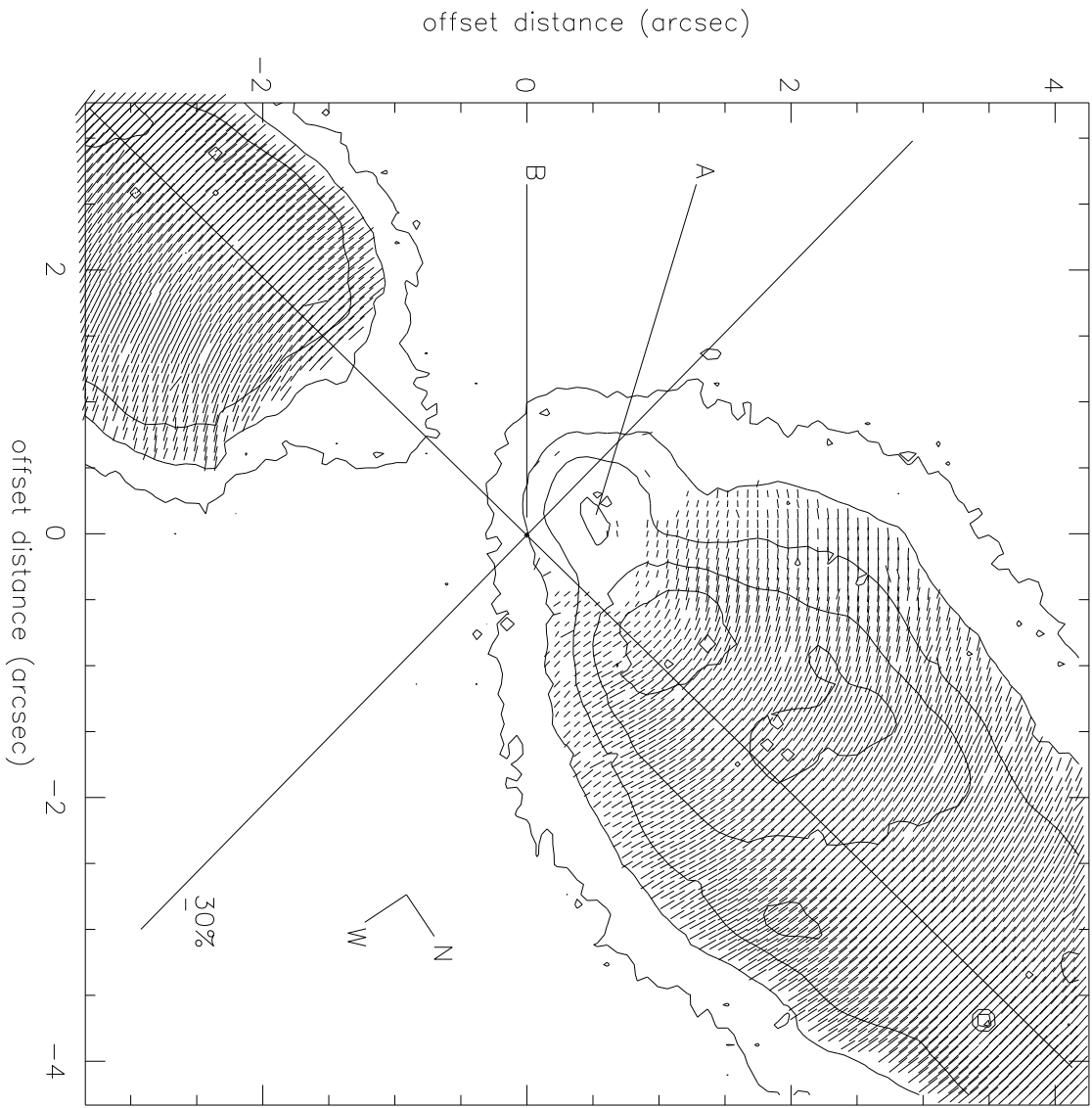
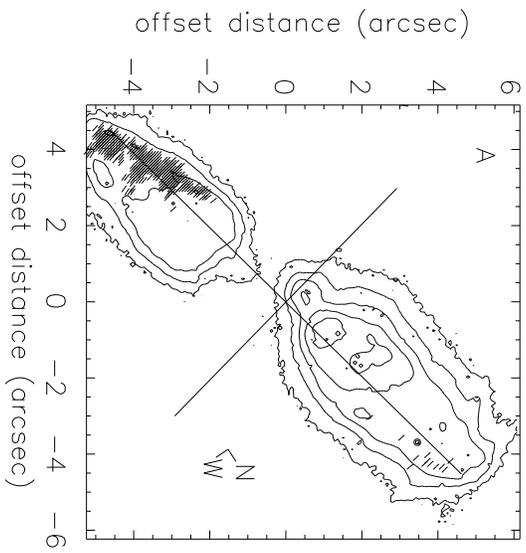
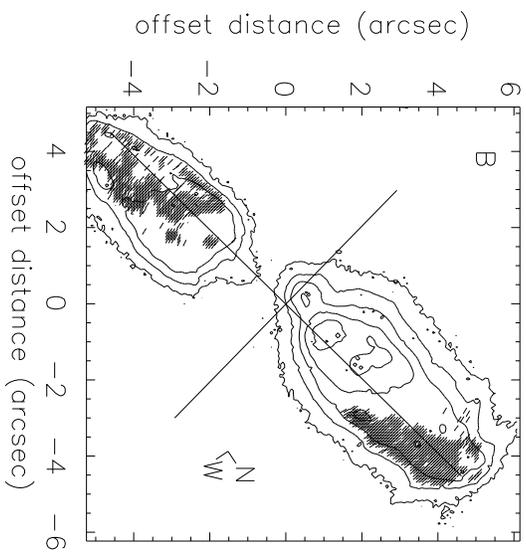
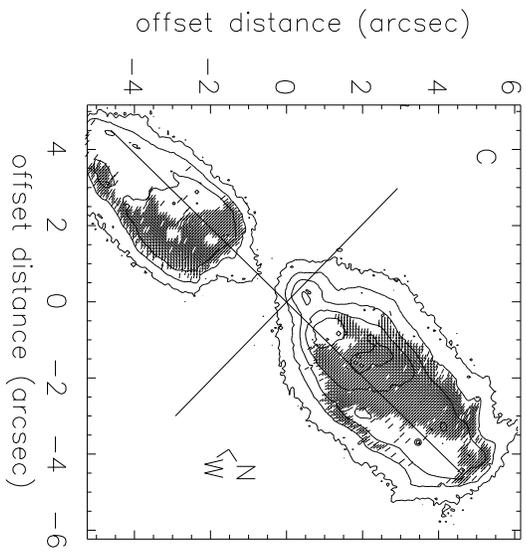
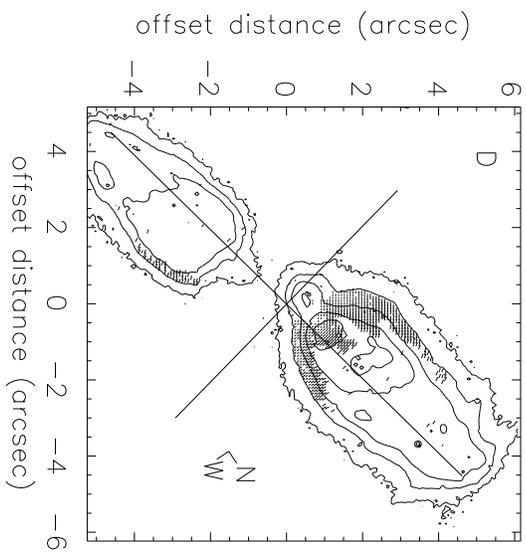
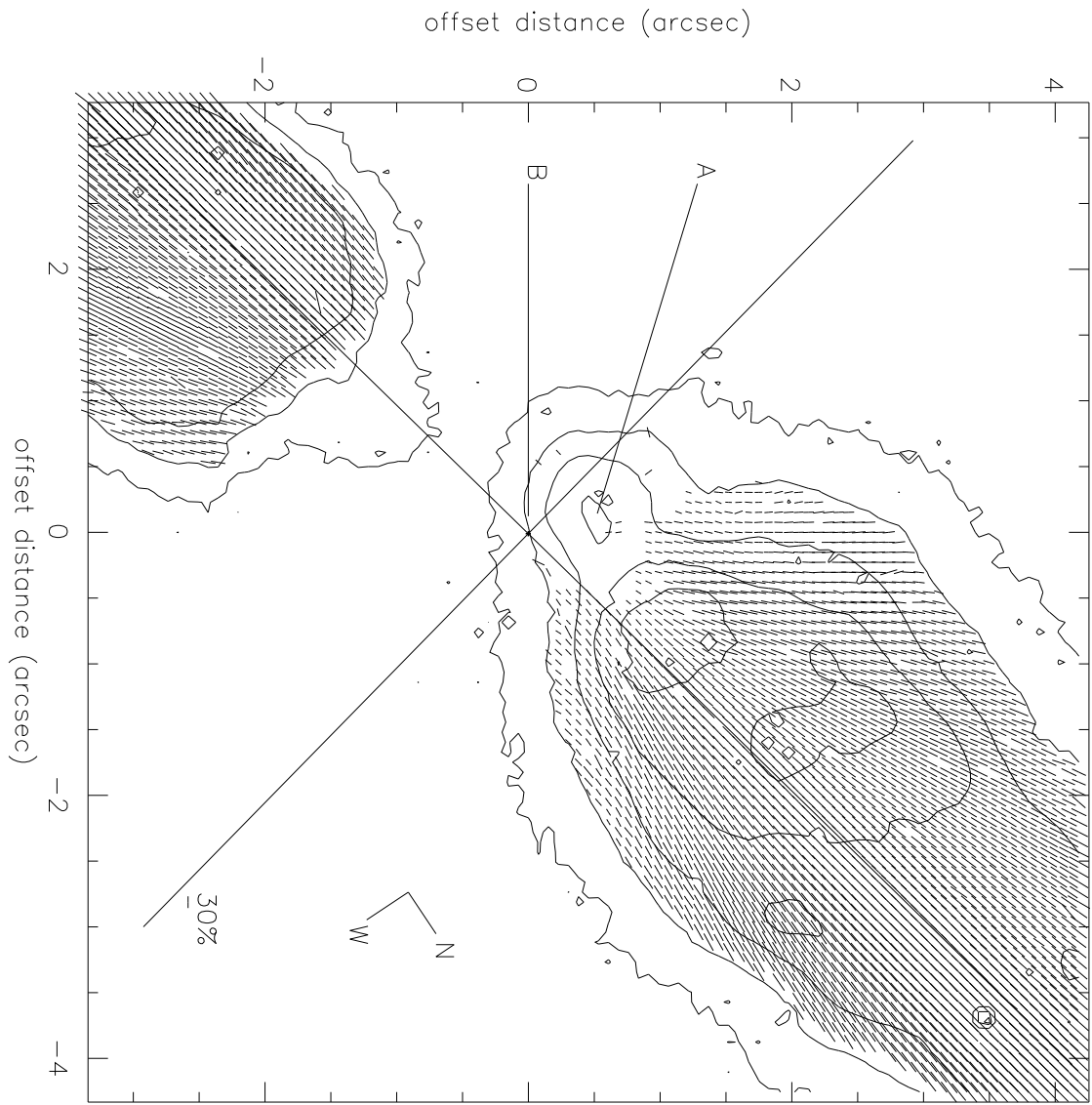
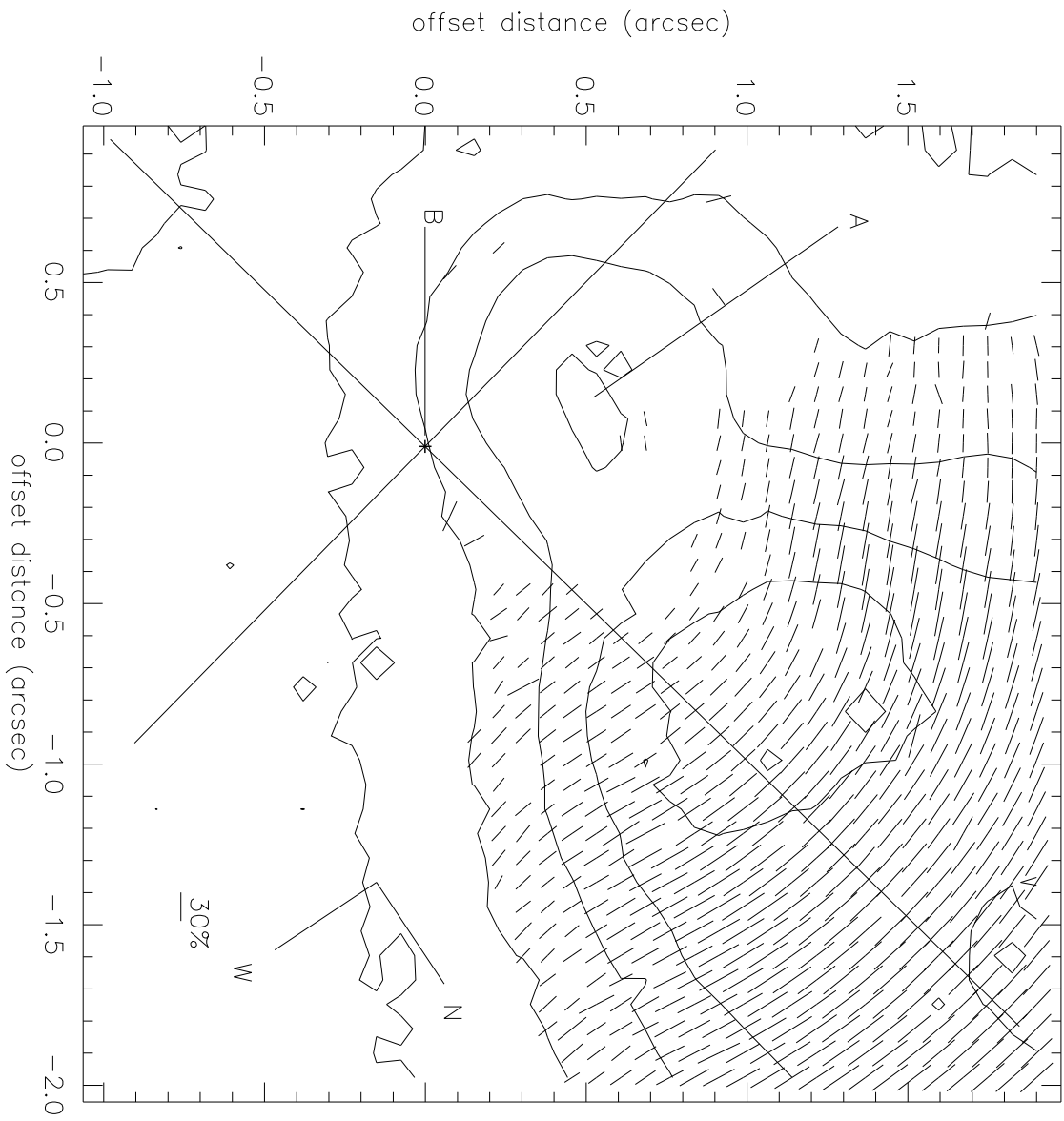
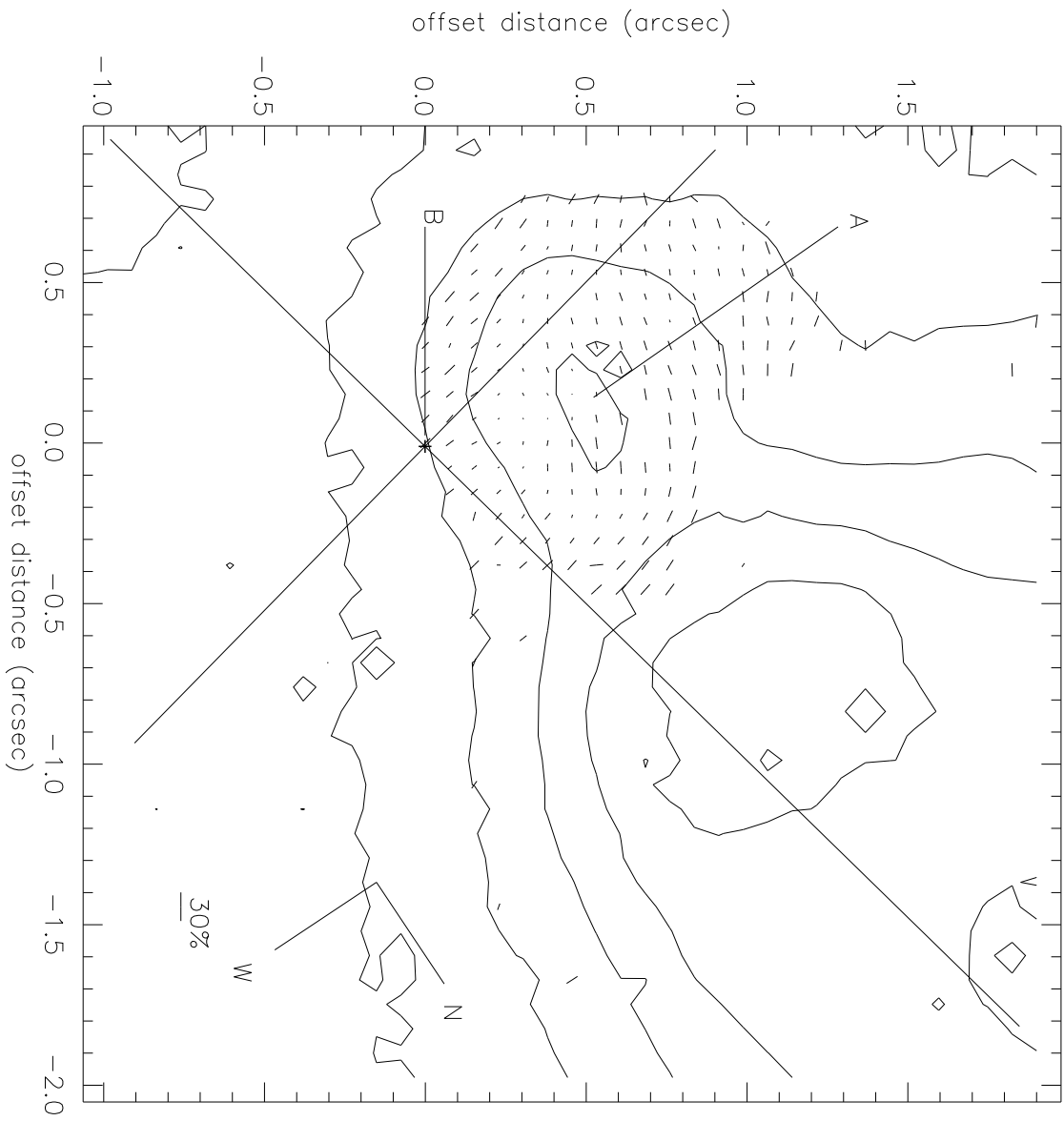
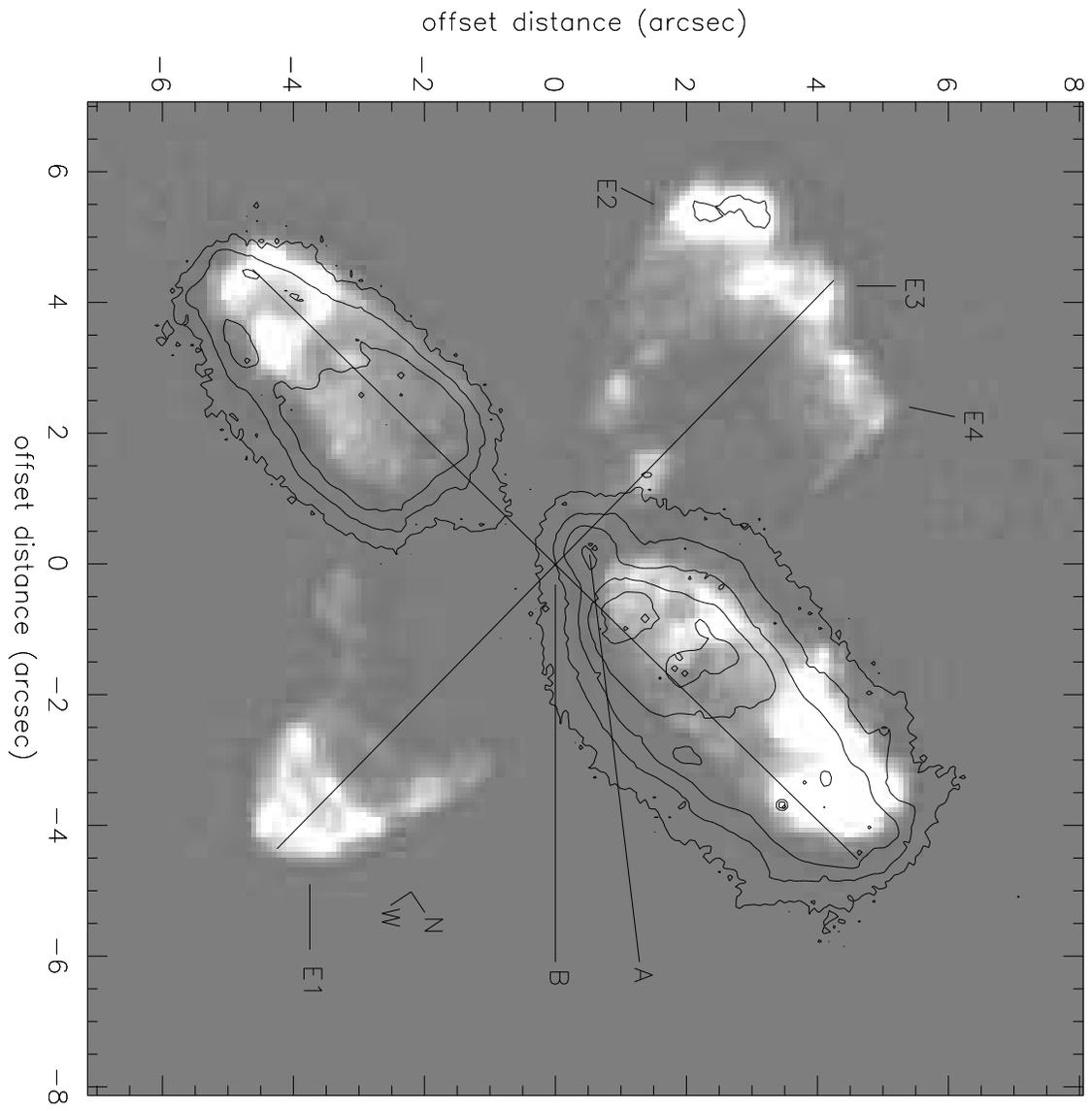
\end{document}